\documentclass[aps,pre,twocolumn,superscriptaddress,floatfix]{revtex4-1}
\usepackage{ifpdf}
 \newif\ifpdf
\ifx\pdfoutput\undefined
   \pdffalse
\else
   \pdfoutput=1
   \pdftrue
\fi
\ifpdf
   \usepackage{graphicx}
\else
   \usepackage{graphicx}
\fi

\usepackage{amsmath,amssymb}

\usepackage{epsfig}

\begin{document}


\title{Superfragile glassy dynamics of onecomponent system with isotropic potential: competition of diffusion and frustration}

\author{R.E. Ryltsev}
\affiliation{Institute of Metallurgy, Ural Division of Russian Academy of Sciences, Yekaterinburg 620017, Russia}

\author{N.M. Chtchelkatchev}
\affiliation{Institute for High Pressure Physics, Russian Academy of Sciences, 142190 Troitsk, Russia}
\affiliation{Moscow Institute of Physics and Technology, 141700 Moscow, Russia}
\affiliation{L.D. Landau Institute for Theoretical Physics, Russian Academy of Sciences, 117940 Moscow, Russia}

\author{V.N. Ryzhov}
\affiliation{Institute for High Pressure Physics, Russian Academy of Sciences, 142190 Troitsk, Russia}
\affiliation{Moscow Institute of Physics and Technology, 141700 Moscow, Russia}


\begin{abstract}
We investigate glassy dynamical properties of one component three-dimensional system of particles interacting via pair repulsive potential by the molecular dynamic simulation in the wide region of densities. The glass state is superfragile and it has high glassforming ability. The glass transition temperature $T_g$ has pronounced minimum at densities where the frustration is maximal.
\end{abstract}

\maketitle

The ubiquitous glass formation and jamming still puzzle physicists~\cite{Berthier,Stillinger}. The microscopic mechanism of the drastic slowing down of the structural relaxation of a liquid on cooling is one of the central issues of the physics of the liquid-glass transition. The question ``why some liquids form a glass easily but others do not,'' is still the matter of debates.

There is a paradigm that one component liquid with isotropic potential typically spontaneously crystallizes being supercooled in (quasi)equilibrium conditions~\cite{Binder_Kob,Kob1999,Molinero_Sastry_Angell,Bernu}. It is a formidable challenge to avoid, e.g., spontaneous crystallization in quasiequilibrium cooling of the onecomponent Lennard-Jones liquid. Yet it was discovered not long ago that there are some exceptions from the paradigm. The common fitch of these exceptions is the pronounced attractive well of the pair potential, see, e.g., \cite{Dzugutov1,Dzugutov2}. Here we show using the molecular dynamics (MD) that the one component simple liquid with \textit{pure repulsive potential} shows glassy behaviour in quasiequilibrium cooling.

Frustration, when one cannot minimize the energy of the  system by merely minimizing all local interactions, is one of the basic factors that stipulates the glass-forming ability~\cite{Berthier,Tanaka}. E.g., it can be related with the long-range alternating interactions (e.g., in spin glasses~\cite{Spin_glass_review}) or with geometrical reasons~\cite{Tarjus-Kivelson}. The potential used in Ref.~\cite{Dzugutov1,Dzugutov2} was optimized to produce icosahedral local order and so geometrical frustration. Our potential has the soft step and the simple liquid with this interaction has two characteristic scales. The competition between these scales makes the system effectively quasibinary~\cite{Ryzhov JCP2008}. And this is the origin of frustration in our system. Intuitively increasing frustration one should favor the formation of the glass. Here we show that on the contrary, the glass transition temperature $T_g$ may have minimum at parameters where the frustration is maximal.

\begin{figure}[b]
  \centering
  \includegraphics[width=0.99\columnwidth]{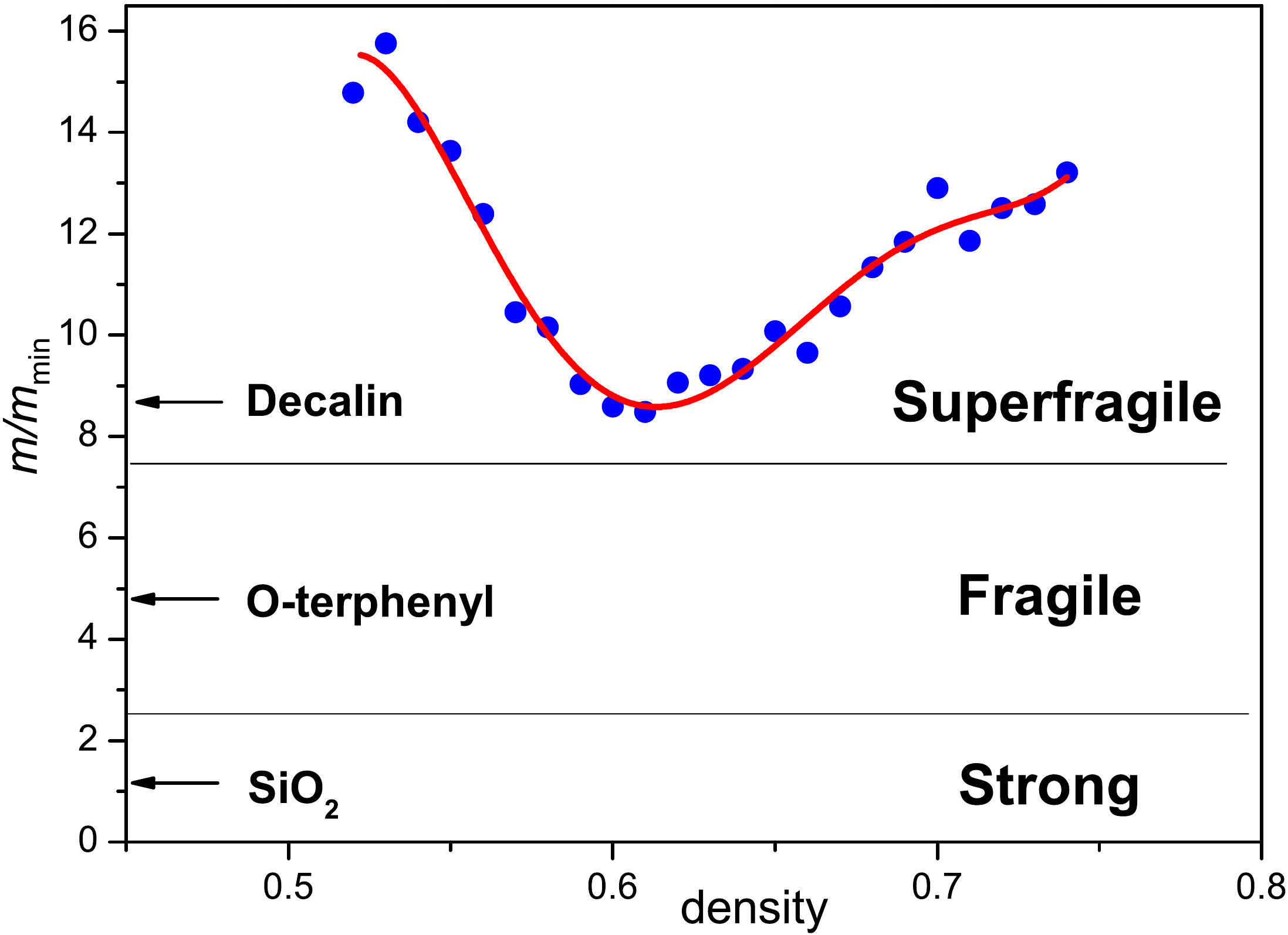}
  \caption{(Color online) Relative fragility of our system as the function of density. For comparison we show the fragility of the typical glassformers. Our system in the glass regime appears to be extremely fragile. It seems that it is in the short list of the most fragile glassy systems.}\label{fragility_comparison}
\end{figure}

The second important concept of glassy physics is fragility. According to Angell-classification~\cite{Angell0,Stillinger} the glassforming liquids effectively divide into two classes: ``strong'' and ``fragile'', where the viscosity of the liquid shows either nearly Arrhenius behaviour with temperature or much faster one. We found the fragility index and concluded that our system is superfragile [its fragility index exceeds that of the ``Decalin'' -- one the most fragile liquid \cite{decalin}], see Fig.~\ref{fragility_comparison}.

We use the pair potential model of ``collapsing soft spheres''~\cite{Ryzhov JCP2008,RyzhovPRE2009,Ryzhov JCP2011_1}:
\begin{gather}\label{csp}
  U(r)=\varepsilon\left(\frac{\sigma}{r}\right)^{n}+\varepsilon n_F\left[2k_0\left(r-\sigma_1 \right)\right],
\end{gather}
where $n_F(x)=1/[1+\exp(x)]$, $\varepsilon$ -- is the unit of energy, $\sigma$ and $\sigma_1$ are ``hard''-core and ``soft''-core diameters. This kind of potentials, Eq.\eqref{csp}, is succesfully used for simulation of water-like anomalies, liquid-liquid phase transitions and glass formation~\cite{Ryzhov JCP2008,RyzhovPRE2009,Ryzhov JCP2011_1,Franzese,Buldyrev-Stanley,Olivera-Barbosa,Angell-buldurev}. The graph of the potential we discuss in the Supplementary and in Fig.~\ref{fig5}.

\begin{figure}[tb]
  \centering
  \includegraphics[width=0.99\columnwidth]{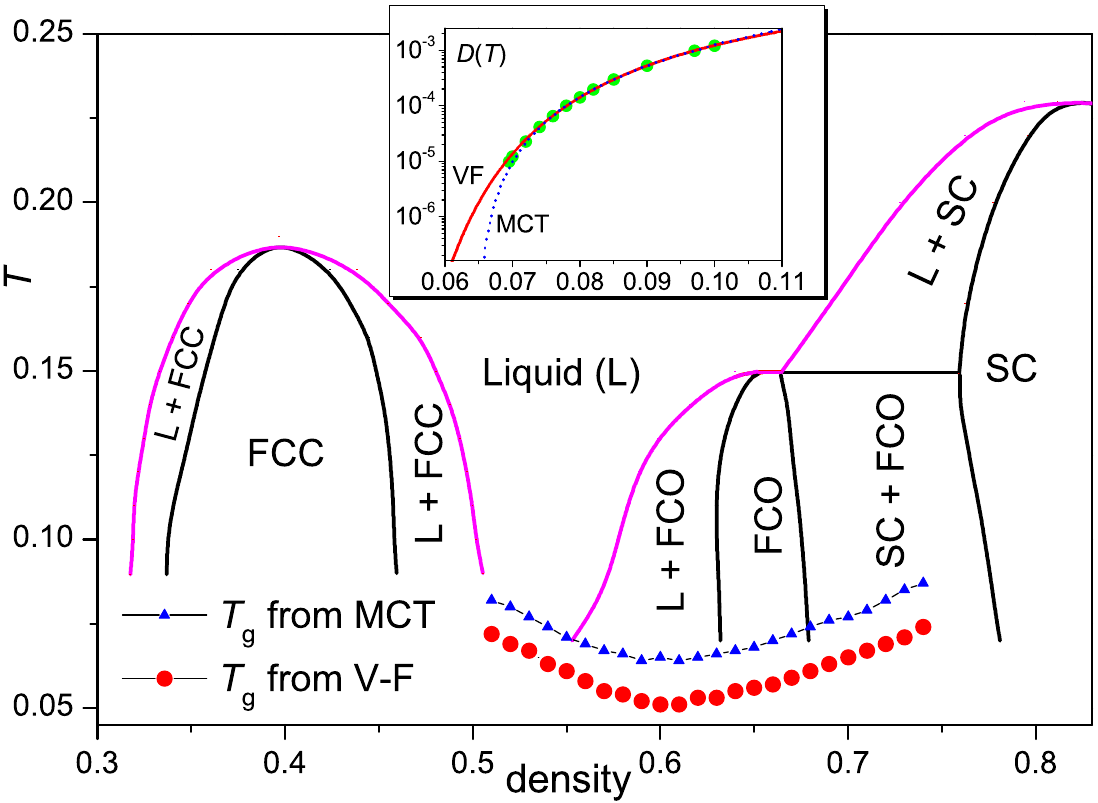}\\
  \caption{(Color online) We show the glass transition temperature on the sketch of the phase diagram obtained in Ref.~\cite{Ryzhov JCP2008}. The red circles and blue triangles correspond to $T_g$ extracted from $D(T)$ using the Vogel-Fulcher formula (VF)~\cite{VF} and the power law from mode-coupling theory (MCT)~\cite{MCT} correspondingly. In the inset we show the accuracy of $D(T)$ approximation by VF and MCT for $\rho=0.6$.  The circles show the result of MD simulations.}\label{phase_diagram}
\end{figure}

In the remainder of this paper we use the dimensionless quantities: $\tilde{{\bf r}}\equiv {\bf r}/\sigma$, $\tilde U=U/\varepsilon$, temperature $\tilde T=T/\epsilon$, density $\tilde{\rho}\equiv N \sigma^{3}/V$, and time $\tilde t=t/[\sigma\sqrt{m/\varepsilon}]$, where $m$ and $V$ are the molecular mass and system volume correspondingly. As we will only use these reduced variables, we omit the tildes. So we take here $n=14$, $k_0=10$, and $\sigma_1=1.35$. These parameters values reveal complex system behavior such as phase diagram with polymorphous transitions and disordered gap, see Fig.~\ref{phase_diagram}, and water-like anomalies~\cite{Ryzhov JCP2008,RyzhovPRE2009,Ryzhov JCP2011_1}.

For MD simulations we have used the system of $N=5000$ particles that were simulated under periodic boundary conditions in Nose-Hoover NVT ensemble. We have checked that $N=5000$ is enough amount of particles to eliminate the finite size effects that agrees with Ref.~\onlinecite{kob}. The MD time step was $\delta t=0.01$. It is nearly the maximum possible time step that satisfies the energy conservation condition. The system was studied in the density region of $\rho\in(0.35-0.75)$. At all densities of this region the system was cooled in a stepwise manner from high temperature state and completely equilibrated at each step until convergence of time dependence of mean square displacement (up to $5\times10^7$ time steps). The time dependencies of temperature, pressure and configuration energy were additionally analyzed to control equilibration. Data were subsequently collected during the time $t_{\rm samp}$ that was chosen to be large enough for correct calculation of diffusion coefficients by Einstein relation. So this time was approximately equal to $t_{\rm samp}\sim 3\tau$, where $\tau$ is the time necessary to reach diffusive regime after ballistic and plateau ones. For more details see Supplementary Material.

\begin{figure}[tb]
  \centering
  \includegraphics[width=0.95\columnwidth]{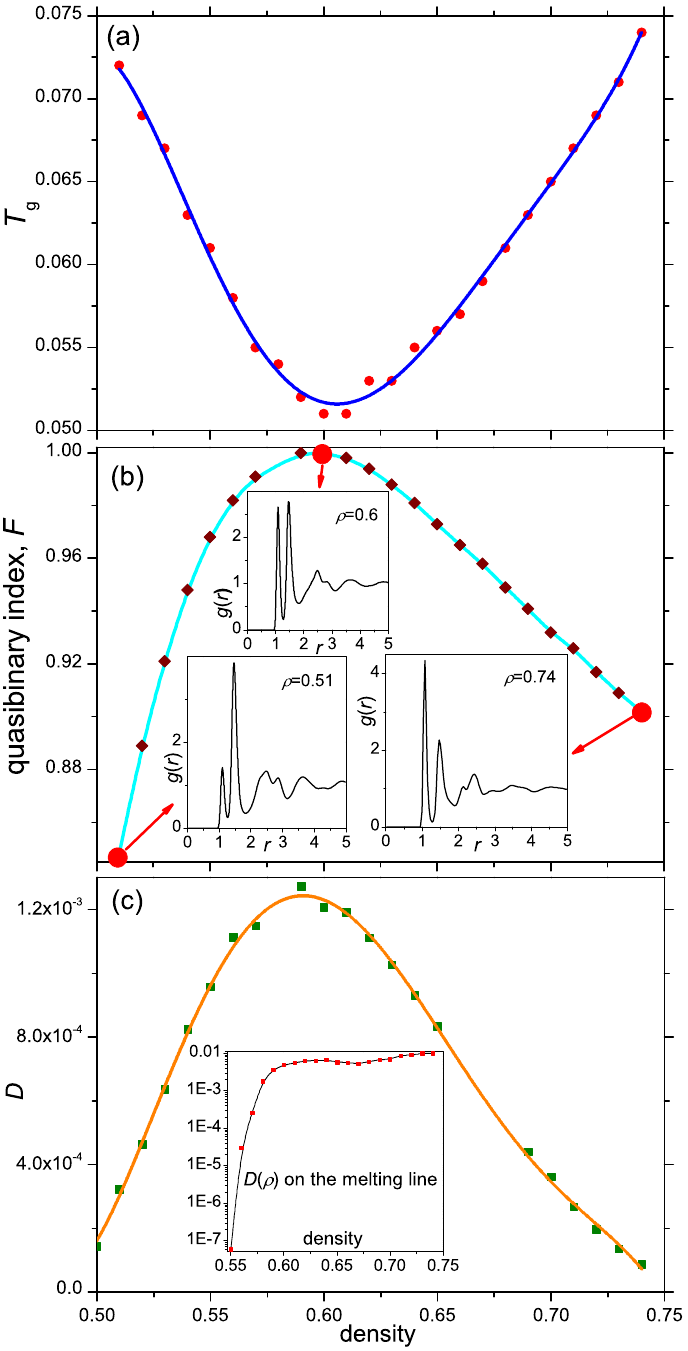}\\
  \caption{(Color online) (a) $T_g(\rho)$ [using (VF)]. (b) Quasibinary index, $F$, for $T=0.1$ and the range of densities where the glassy dynamics was detected. Comparing the $F(\rho)$  with $T_g(\rho)$  we see that these dependencies are opposite to one another: the bigger $F$ the smaller $T_g$ and the maximum of the former curve is located at the same density as the minimum of the later one. The inset shows the radial distribution function $g(r)$. It has two clear peaks at $r=d$, $r=\sigma$ that proves the quasibinary behaviour of the system. (c) The diffusion coefficient $D(\rho)$ for $T=0.1$.}\label{Tg_F_D}
\end{figure}

Avoiding spontaneous crystallization during equilibration process is the principal difficulty of MD simulations of glassy dynamics of liquids. For model glass-forming systems this problem is usually solved by either using of non-isotropic potentials~\cite{Molinero_Sastry_Angell,Beest}, or considering multicomponent systems~\cite{Weber,Kob}, or using nonequilibrium cooling~\cite{Buldyrev-Stanley,Xu-Buldyrev-Giovambattista-Angell-Stanley}. For collapsing soft spheres system, it is possible to avoid crystallization in the one-component system with isotropic potential due to the quasibinary behavior that develops itself in certain density interval. In our case this range is $\rho\in(0.51-0.74)$. In  the inset in Fig.~\ref{Tg_F_D}b we show the splitting of the first peak in the radial distribution functions (RDF) that illustrates the quasibinary behaviour of our system in hand. Outside this interval, it is hardly possible to study the glassy dynamics because system spontaneously crystalizes whereupon supercooling below the melting line. Conversely, inside the region mentioned, one can equilibrate supercooled liquid without crystallization down to temperatures at which relaxation time becomes too large for simulation. In our case, these minimal temperatures were chosen so that diffusion coefficients were on the order of $10^{-5}$. At that temperatures, the total calculation time required for equilibration and data collecting is $\sim3\times 10^7$ MD steps [three days of calculation on 32 processors in parallel].

In order to control stability of glassy state, we performed calculation at lower temperatures also (down to $T_g$). At that temperatures, the system cannot be equilibrated completely because of large relaxation times and so we did not use these data for $T_g$ calculations. But what we have observed is the absence of any crystallization up to $10^8$ MD steps and so we conclude that the glass state is stable (at least at simulation time scales).

\begin{figure}[tb]
  \centering
  \includegraphics[width=0.95\columnwidth]{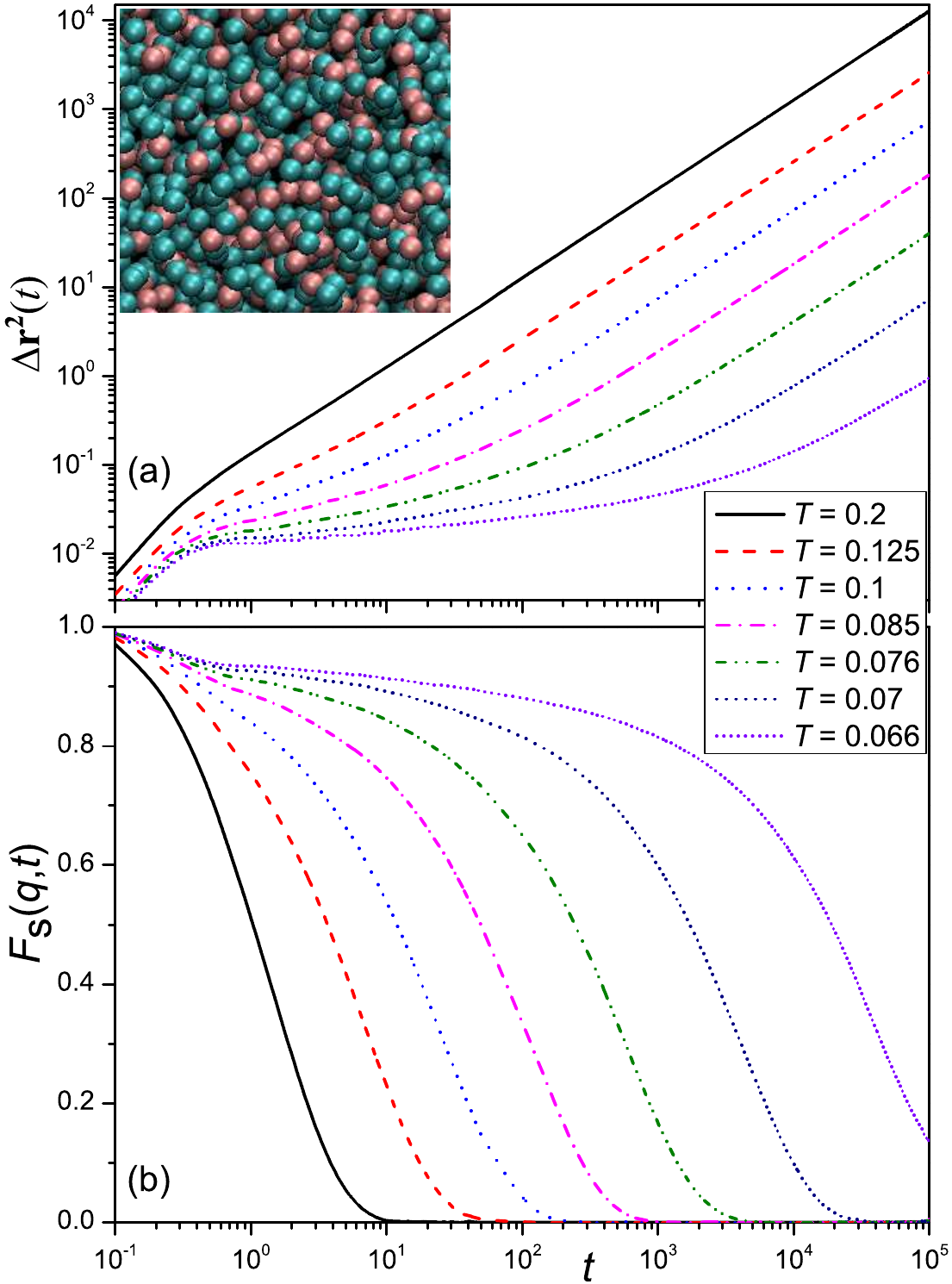}
  \caption{(Color online) (a) Mean-square displacement, and (b) Intermediate scattering function for the density $\rho=0.6$. Inset shows the snapshot of the typical distribution of particles for $T=0.1$ and $\rho=0.6$. We mark by red (blue) color particles that have a neighbor at distance of the order of the first (second) maximum in rdf [hard-core and soft-core diameters].}\label{dr_Fs}
\end{figure}
To acquire information about the glass state we focus on the temperature range $T\gtrsim T_g$ where the ``glassforming fluctuations''\cite{chtchTMF} slow down the system dynamics with temperature decreasing. The conventional correlation function tools have been used: the mean square displacement $\Delta r^2(t)$ and the intermediate scattering function $F_s(q,t)$. The time dependencies of these functions for $\rho=0.6$ and different temperatures are shown in Fig.~\ref{dr_Fs}. One can see the typical picture glassformers demonstrate at low temperature~\cite{glass_book2012}. Namely, the ``plateau'' reflecting the cage effect (when the particle is trapped in the ``cage'' of the nearest neighbours) appears on $\Delta r^2(t)$ and $F_s(q,t)$ at sufficiently low temperatures that indicates the onset of glassy regime in system dynamics. Meanwhile, the system remains in disordered state as can be seen from radial distribution functions, see inset of Fig.~\ref{Tg_F_D}b. We note the splitting of the first and second peaks of the radial distribution function (Fig.~\ref{Tg_F_D}b). The splitting of the first peak reflects the quasibinarity of the system caused by the form of the potential (see discussion below). While the splitting of the second peak is apparently the (system independent) attribute of glassy state~\cite{Dzugutov1,Buldyrev-Stanley,Mizuguchi}.

In order to estimate the glass transition temperature, we calculated diffusion coefficient $D$ at each of the temperature and density investigated. The temperature dependencies $D(T)$ were approximated by both the Vogel-Fulcher formula (VF)~\cite{VF,footnote}
\begin{equation}\label{Vogel-Fulcher law}
D=D^{\rm (vf)}_0\exp\left(-\frac{A}{T-T^{\rm (vf)}_0}\right)
\end{equation}
and the power law from mode-coupling theory (MCT)~\cite{MCT,footnote}
\begin{equation}\label{D MCT}
D=D^{\rm (mc)}_0 \left({T-T^{\rm (mc)}_0}\right)^\gamma.
\end{equation}
The parameters $D^{\rm (vf)}_0$, $A$, $T^{\rm (vf)}_0$, $D^{\rm (mc)}_0$, $T^{\rm (mc)}_0$, $\gamma$ were obtained using method of nonlinear least squares. Since the expressions (\ref{Vogel-Fulcher law}), (\ref{D MCT}) are correct only in a vicinity of glass transition temperature, the temperature interval $T\in (T_{\rm min},\, T_{\rm max})$ for least squares approximation was chosen so that $D(T_{\rm min})\simeq 10^{-5}$, $D(T_{\rm max})\simeq 10^{-3}$ that approximately corresponds $T/T_g\in (1.15, 1.6)$. The typical temperature dependence $D(T)$ of diffusion coefficient obtained from simulation and its VF and MCT approximations are shown in the inset of Fig.~\ref{phase_diagram}. One can see that both formulas provide good fitting of simulation data.

Having the parameters of (\ref{Vogel-Fulcher law}), (\ref{D MCT}) one can get the glass transition temperature. According to generally accepted definition, glass transition occurs at $D_0/D=10^n$, where $13\lesssim n\lesssim17$. Using (\ref{Vogel-Fulcher law}), (\ref{D MCT}) we obtain
\begin{gather}\label{Tg V-F}
T_g^{\rm (vf)}=T^{\rm (vf)}_0+\frac{A\log_{10} e}{n},\qquad
\\\label{Tg MCT}
T_g^{\rm (mc)}=T^{\rm (mc)}_0+10^{-n/\gamma}.
\end{gather}

The graph of $T_g(\rho)$ is shown in Fig.~\ref{phase_diagram} on the phase diagram (previously obtained in \cite{Ryzhov JCP2008}). It follows that $T_g(\rho)$ dependencies obtained using VF and MCT approaches are in good agreement with each other. Particularly, the both curves are located under the melting line and have a minima in the vicinity of density $\rho=0.6$.

\begin{figure}[tb]
  \centering
  \includegraphics[width=0.99\columnwidth]{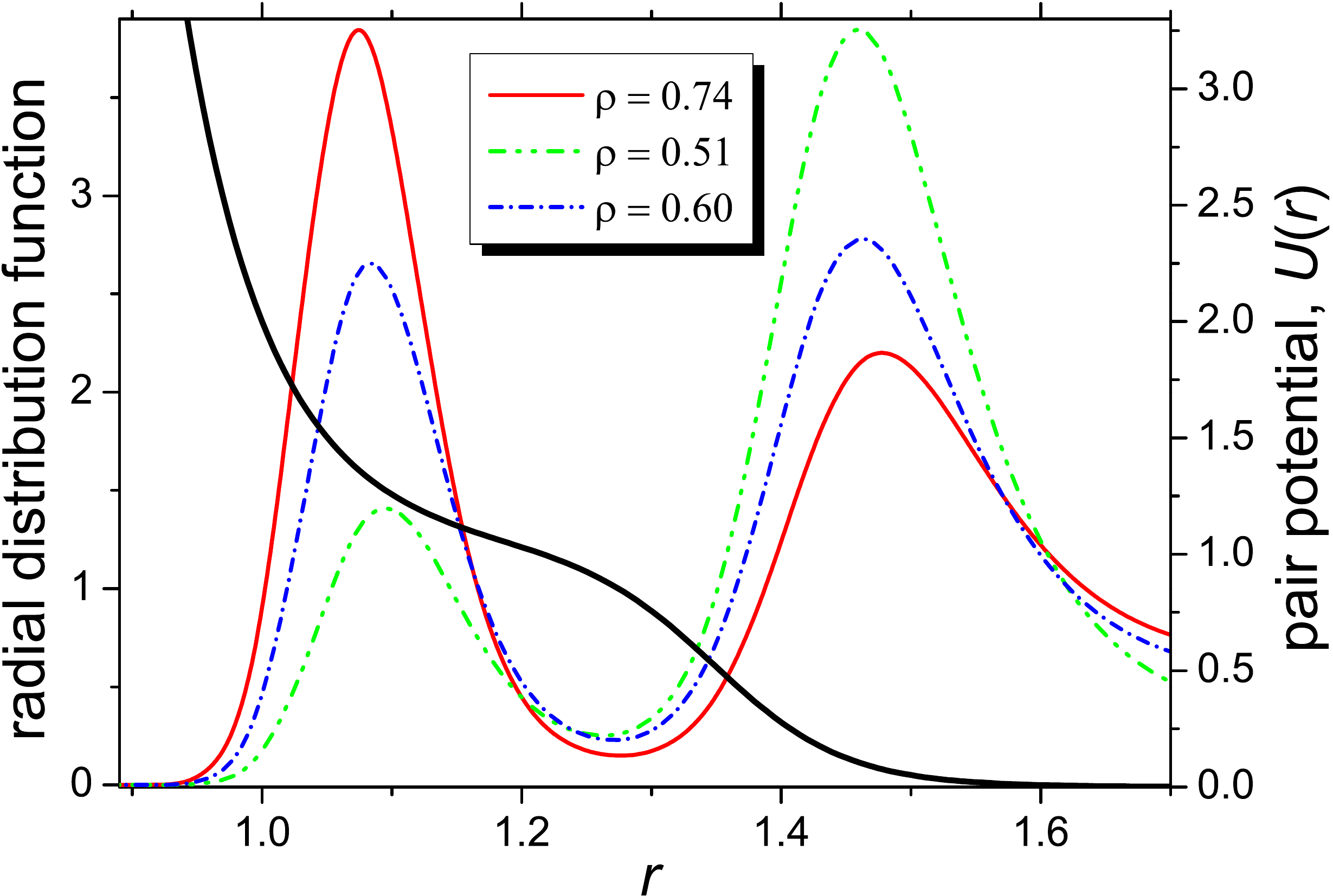}
  \caption{(Color online) The pair potential $U(r)$ and the splitted first peak of the radial distribution functions.  }\label{fig5}
\end{figure}
We calculate the fragility index $m$ showing the deviation of $D(T)$ from Arrhenius law and allowing to identify the type of glass-forming system according to strong-fragile classification. Starting with the definition of fragility in the form~\cite{Bohmer,Shi,Coniglio,Sengupta}
\begin{equation}\label{fragility def}
m = \left. {\frac{{\partial \log _{10} \left( {D_0 /D} \right)}}{{\partial \left( {T_g /T} \right)}}} \right|_{T = T_g }
\end{equation}
and using Eqs.~\eqref{Vogel-Fulcher law},\eqref{Tg V-F} we get
\begin{equation}\label{Fragility V-F}
m = n\left( {\frac{{T_0^{({\rm vf})} }}{A}n\ln 10 + 1} \right).
\end{equation}
The equation (\ref{Fragility V-F}) particularly shows that the fragility index has minimal value $m_{\rm min}=n$ corresponding to the limit ${T_0^{({\rm vf})} }/{A}\rightarrow 0$ that turns (\ref{Vogel-Fulcher law}) to Arrhenius law. Fig.~\ref{fragility_comparison} shows the density dependence of the reduced fragility, $m/m_{\rm min}$, of our system in comparison to several glassformers. One can see that the dynamics of our system is extremely fragile since the mean value of its fragility exceeds the one for decalin --- one of the most fragile system \cite{decalin}. Note that $m(\rho)$  is nonmonotonic and reveals clear minimum at $\rho=0.6$ as well as for $T_g(\rho)$, see Fig.~\ref{phase_diagram}.
Recently it was shown that the increase of the interaction softness can lead to the increase of the fragility~\cite{Shi,Sengupta}. The ``softness'' of our pair potential is large at $r$ corresponding to the peaks of the radial distribution function, see Fig.~\ref{fig5}. Away from $\rho^*=0.6$ one of the RDF peaks dominates, see inset in Fig.~\ref{Tg_F_D}b, that effectively makes softer the effective interparticle interaction and helps to interpret the nonmonotonic behaviour of $m(\rho)$.

Advances over the last decade have linked non-Arrhenius behavior of fragile glass formers with the presence of locally heterogeneous dynamics: i.e. the presence of distinct (if transient) slow and fast regions within the material~\cite{Stillinger,Angell2000}. The quasibinary character of our particle system allows to make the selection among the particles: we mark by red (blue) color particles that have a neighbor at distance of the order of the first (second) maximum in rdf [hard-core and soft-core diameters], see Fig.\ref{fig5}. The snapshot of the spatial particle distribution over the simulation volume (see inset in Fig.~\ref{dr_Fs} and Fig.~2 in the Supplementary) shows the high degree of the heterogeneity in our system that favors the existence of the locally heterogeneous dynamics within the system because of different free volumes (and so diffusion coefficients) of particles with different effective radii.

It has been mentioned above that the system demonstrates quasibinary behavior due to the repulsive shoulder of the pair potential, see inset in Fig.~\ref{Tg_F_D}b.  It reflects the competition between hard-core and soft-core scales and, as a result, the frustration in the system. As the quasibinarity index we choose
\begin{equation}\label{Frustration index}
F=\frac{2\rho_1\rho_2}{\rho_1^2+\rho_2^2},
\end{equation}
where $\rho _1  = \frac{{\rho \int_0^{r_1 } {r^2 g(r)dr} }}{{\int_0^{r_1 } {r^2 dr} }}$ and $\rho _2  = \frac{{\rho \int_{r_1 }^{r_2 } {r^2 g(r)dr} }}{{\int_{r_1 }^{r_2 } {r^2 dr} }}$. Here $r_1$ and $r_2$ are minima of the first and the second RDF peaks. So $\rho_1$, $\rho_2$ are local densities in the vicinity of the peaks and $F$ is their inverse symmetrized ratio.

The origin of the frustration in our system is the same as for binary soft-sphere like systems where the frustration is due to the large, local rearrangement of atoms required for the formation of a crystal from a fluid or glassy configuration~\cite{about_F}. This situation is caused by presence of second component and so the frustration of such type increases with increasing concentration. Thus the quasibinary index $F$ can serve as frustration measure in our system. It is clear that the value $F\simeq 0$ corresponds to the situation than only one of the scales dominates and so there is no frustration in the system. On the contrary, if $F\simeq 1$ the competition between hard-core and soft-core scales and so the frustration are maximal. It should be noted that $F$ depends on temperature as well as on density. In order to study the influence of frustration on glass-forming ability of the system we calculate the density dependence of the frustration index at sufficiently low temperatures. In Fig.~\ref{Tg_F_D}b we show $F(\rho)$ at $T=0.08$ (of the order of $T_g$ for all the densities investigated). It follows from Fig.~\ref{Tg_F_D} that $F(\rho)$ and $T_g(\rho)$ have opposite behaviour: the bigger $F$ the smaller $T_g$ and the maximum of the former curve is located at the same density as the minimum of the later one.

At the density $\rho^*$ where $T_g(\rho)$ has minimum the system has potentially a number of different lattice constants as we mentioned above. If we imagined the long range order formation at $\rho^*$ then the crystal would be strongly distorted by defects due to the competition of the different lattice constants. This situation should favour the diffusion and frustration at the same time, see Fig.~\ref{Tg_F_D}. $T_g$ is determined both by the diffusion and by the frustration: diffusion tries to decrease $T_g$ while the frustration does the opposite. However the diffusion defeats frustration in our system, so $T_g$ has minimum at $\rho^*$.

Finally we discuss the boundaries of the density domain, $\rho\in(0.35-0.75)$. Beyond these boundaries the glass forming ability quickly decreases. This one can judge from the quick rise of the diffusion coefficient, $D(\rho)$, at the melting line, see inset in Fig.~\ref{Tg_F_D}c, and the destruction of the quasibinary behaviour, see inset in Fig.~\ref{Tg_F_D}b. 

In conclusion, we show using the molecular dynamics that the one component simple liquid with pure repulsive potential shows glassy behaviour in quasiequilibrium cooling. We explain the nonmonotonic density dependence of $T_g$, frustration, the diffusion coefficient and fragility by the evolution of the quasibinary properties of our system. We observe that our system belongs to the short list of the most fragile systems.  Our system can be used as the simple toy model for investigation of the quasibinary frustrated systems. In the search for (super)fragility in the one component repulsive simple liquids one should test them for quasibinarity.

We thank V.~Brazhkin, Yu.~Fomin, G.~Rusakov, E.~E.~Tareyeva, E.~Tsiok, L.~D.~Son and M.~Vasin for helpful discussions. The work was supported by Russian Foundation for Basic Research (grants  12-03-00757-a, 10-02-00882-a, 10-02-00694a, 10-02-00700 and 11-02-00-341a), Ural Division of Russian Academy of Sciences (grant RCP-12-P3) and Presidium of Russian Academy of Sciences (program № 12-P-3-1013). We are grateful to supercomputer center of Ural Branch of Russian Academy of Sciences for the access to ``Uran'' cluster.

\appendix
\section*{\large Supplementary Material}

\section{\label{sec:sim}Details of simulation and results handling}

\subsection{MD simulation}
\begin{figure}[b]
  \centering
  \includegraphics[width=0.80\columnwidth]{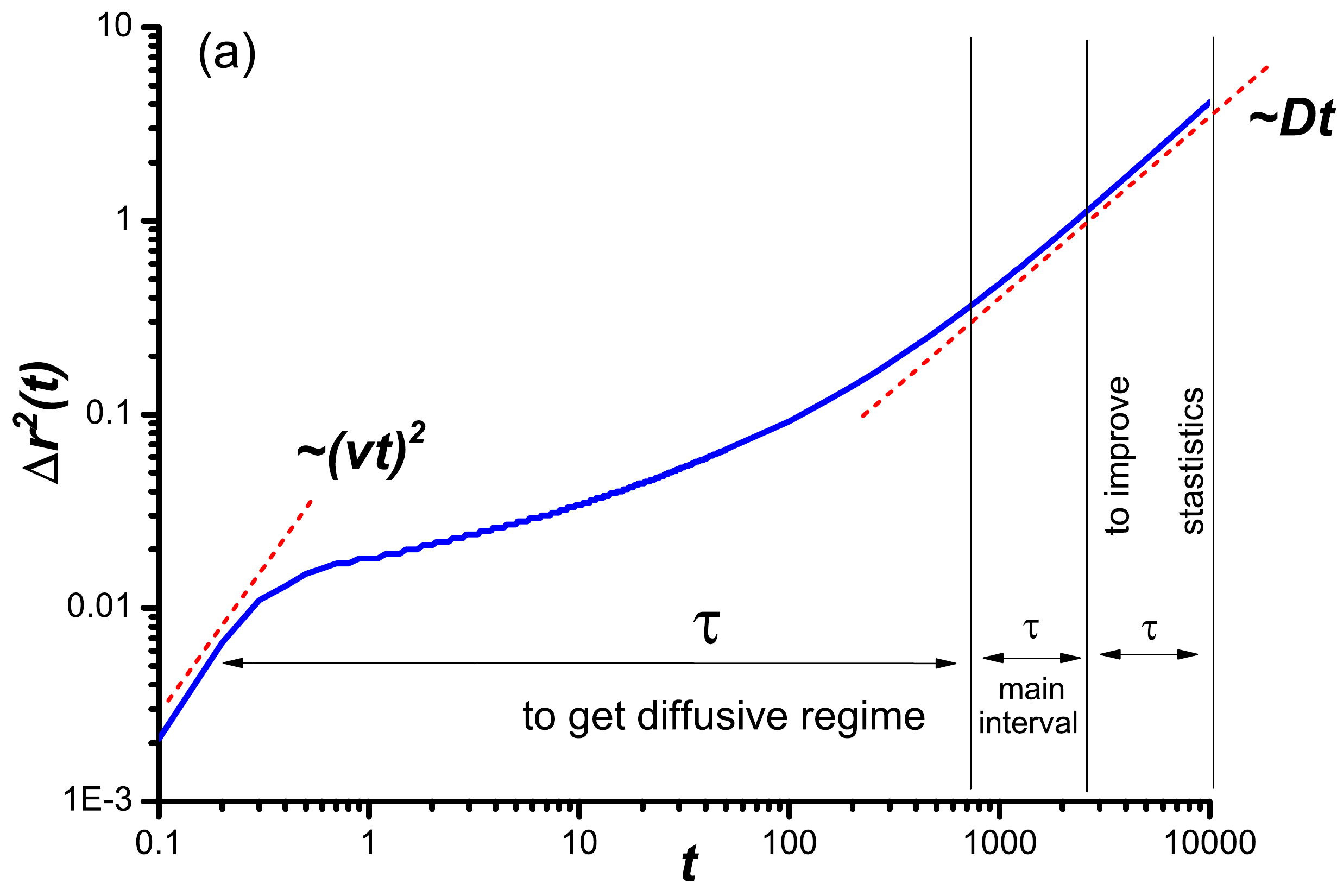} \includegraphics[width=0.80\columnwidth]{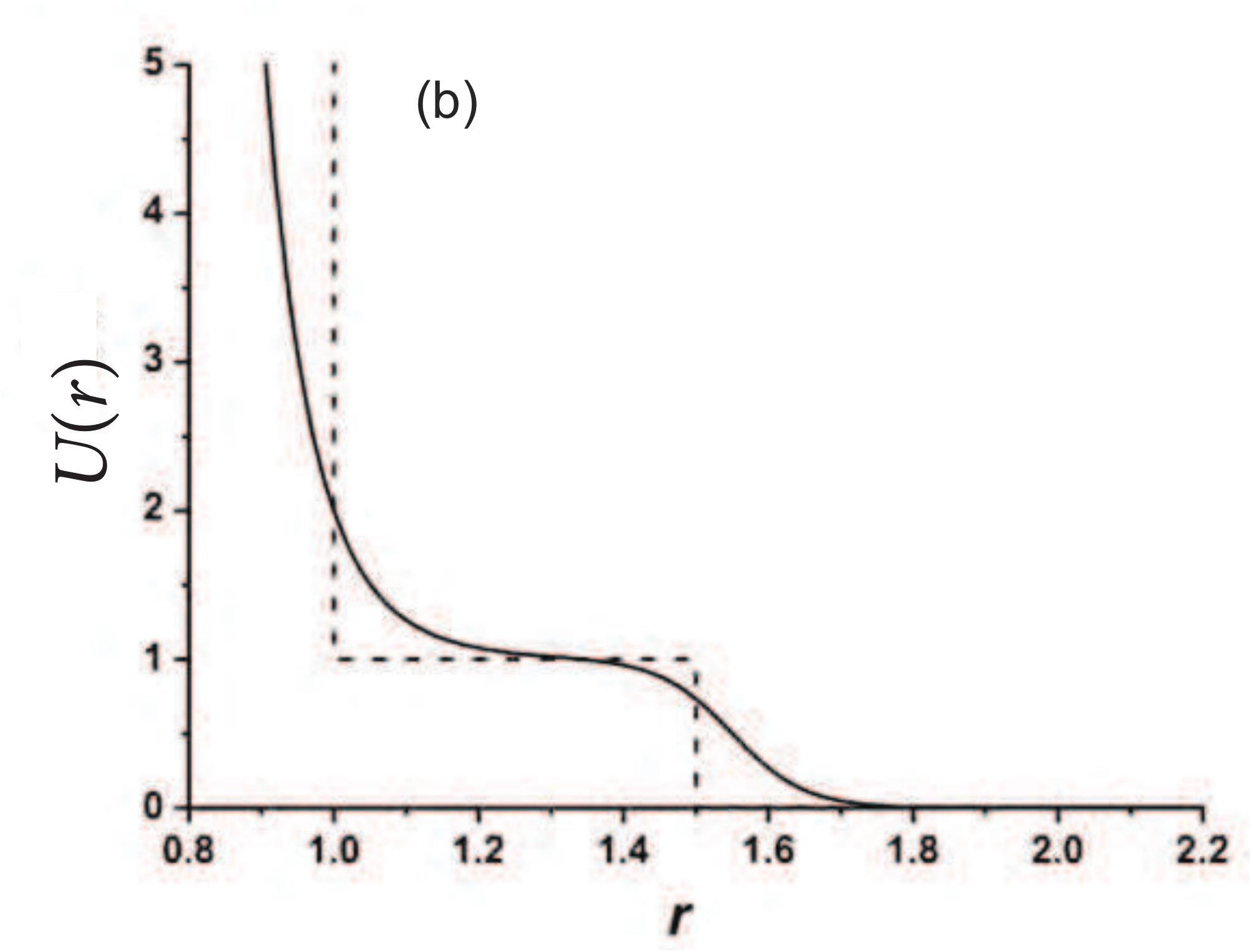} \includegraphics[width=0.80\columnwidth]{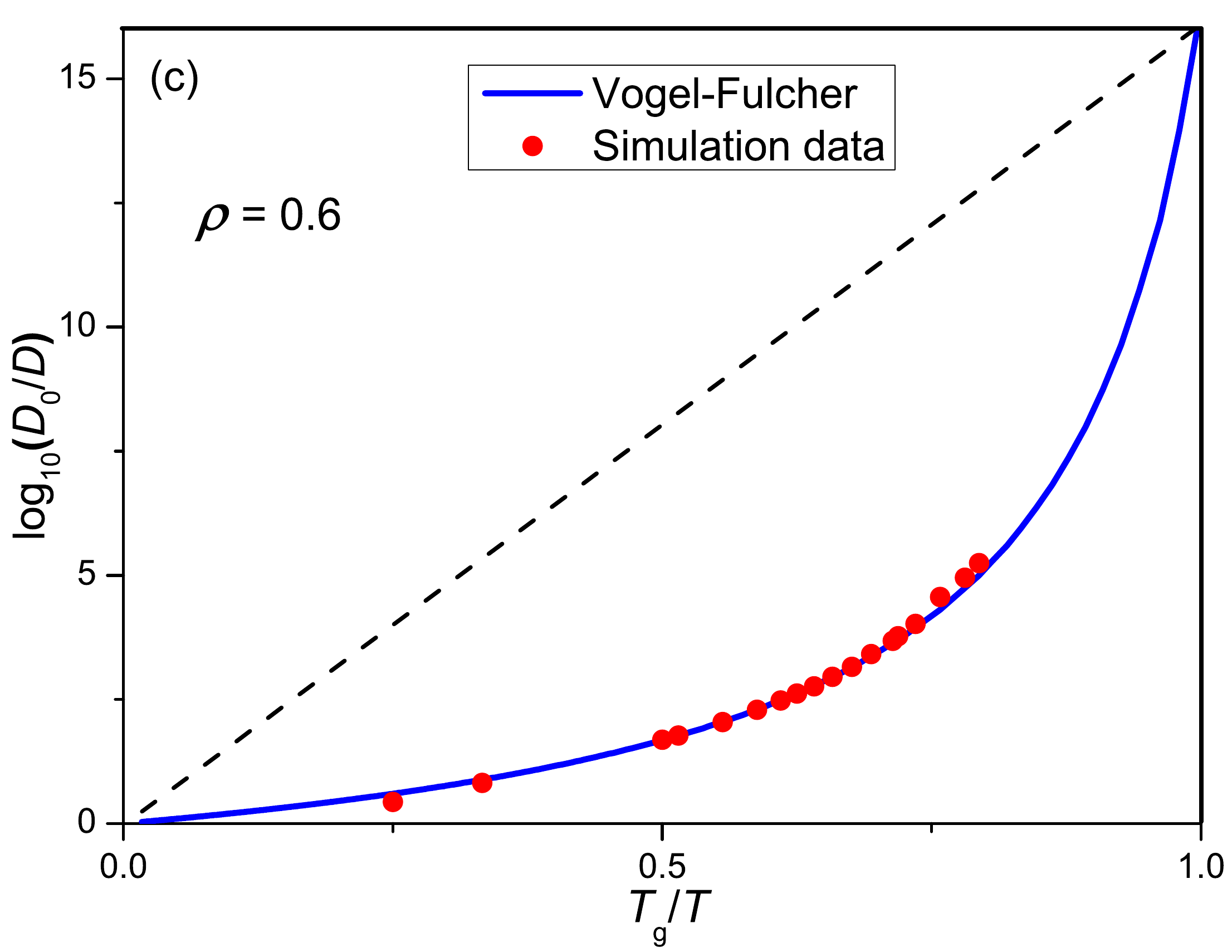}
  \caption{(Color online) (a)``$3\tau$ rule'' for correct calculation of diffusion coefficients from mean square displacement. This rule says that the total time for data collecting should be in order of $3\tau$, where $\tau$ is the relaxation time. The reason is simple: one $\tau$ is needed to reach diffusive regime after ballistic and plateau ones, another $\tau$ is the main time interval for calculation by Einstein formula and one more $\tau$ is to improve averaging statistics for MSD calculations in the main interval. (b) The potential of the collapsing soft spheres model. (c) Angel plot for temperature dependence of diffusion coefficient demonstrating essential deviation of $D(T)$ from Arrhenius law and so the fragile character of system dynamics.}\label{3taurule}
\end{figure}

\textit{For MD simulations}, we have used $\rm{DL\_POLY}$ 2.20 Molecular Simulation Package~\cite{dlpoly} developed at Daresbury Laboratory.

Initially, the system was equilibrated from  lattice configuration at high temperature for $10^6$ MD time steps and then it was consecutively equilibrated at lower temperatures from final configurations of previous runs. At relatively high temperatures ($T \gtrsim 1.2T_g$), the equilibration time was chosen to be in order of $10^4-2\times10^6$ time steps. The time dependencies of temperature, pressure and configuration energy were analyzed to control equilibration. At low ($T < 1.2T_g$) temperatures addition control is needed so systems was consecutively equilibrated until convergence of time dependence of mean square displacement.  Data were subsequently collected during the time $t_{\rm samp}$ that was chosen to be large enough for correct calculation of diffusion coefficients by Einstein relation. So this time was approximately equal to $t_{\rm samp}\sim 3\tau$, where $\tau$ is the relaxation time, see Fig.~\ref{3taurule}(a). The reason is simple: one $\tau$ is needed to reach diffusive regime after ballistic and plateau ones, another $\tau$ is the main time interval for calculation by Einstein formula and one more $\tau$ is to improve averaging statistics for MSD calculations in the main interval.

We obtain $T_g$ by quasiequilibrium method that is just approximation of equilibrium $D(T)$ curve for supercooled liquid by appropriate fitting formula (for example, Vogel-Fulcher law) and using the relation $D(T_g)/D_0=10^n$, where $n=13-17$.  This method is natural for (quasi)equilibrium simulation we have used but it differs from the dynamic method. According to the dynamic method, glass transition temperature is obtained as temperature at which the system falls out of equilibrium (i.e. when the cooling rate exceeds the relaxation one). This temperature can be determined by sharp peaks on heat capacity $C_p(T)$ or thermal expansion coefficient, $\alpha_p(T)$. The temperature, $T_g$, obtained by the dynamic method, depends on the cooling rate $v$, i.e. $T_g=T_g(v)$. The natural question appears: how do the glass transition temperatures obtained by static (quasiequilibrium) and dynamic methods correlate with each other? The answer is not conclusive because the rigorous theory is still absent but some arguments may be given. There are a number of experimental and theoretical works (see, e.g., Refs.~\cite{Samwer,KobTgv,Kob1996,Kob1999,Stillinger,Xu-Buldyrev-Giovambattista-Angell-Stanley,Evenson,Vasin2011}) where $T_g(v)$ dependence has been investigated. It was shown that this dependence can be represented in the form,
\begin{gather}
 T_g(v) = T_g(v=0) + f(v),
\end{gather}
where $f(v)$ is some function of the cooling rate $v=\frac{dT}{dt}$, where $T$ and $t$ are temperature and time correspondingly. The function $f(v)$ is often approximated as follows:
\begin{gather}\label{eqfv}
 f(v)=\frac{B}{\ln\left(v_0/v\right)},
\end{gather}
where $B$ and $v_0$ are constants~\cite{Samwer,Kob1996}. Expression~\eqref{eqfv} is able to give a satisfactory fit to the experimental data when $v$ is varied
over 3~decades~\cite{Samwer}. The shape of Eq.~\eqref{eqfv} can be justified if we use the Vogel-Fulcher approximation for the relaxation time,
\begin{gather}\label{tauVV}
  \tau(T)=\tau_0\exp\left[\frac B{T-T_g(v=0)}\right],
\end{gather}
and if we assume [see, e.g., Refs.~\cite{Samwer,Kob1996,Vasin2011}] that the system falls out of equilibrium at temperature $T_g(v)$ where the relaxation time is of the order of the inverse of the cooling rate, i.e. $T_g(v)/\tau[T_g(v)]\sim v$ [and so $v_0\sim T_g(v=0)/\tau_0$]. It follows from the derivation that Eq.~\eqref{eqfv} is applicable while $v\lesssim v_0$, see, e.g., Ref.~\cite{Vasin2011}. Note that there are other suggestions for $f(v)$ that also agree not so bad with experiments~\cite{Samwer,KobTgv,Kob1996,Vasin2011}. In any case, at sufficiently small cooling rate $T_g$ becomes nearly independent on the cooling rate $v$. This is the limit where we perform our simulations. Then pressure and temperature are well defined in Nozier-Hover thermostat that we have used.

\subsection{The potential of the collapsing soft spheres model}

The potential of the \textit{collapsing soft spheres} model is the pure repulsive potential with two characteristic scales corresponding to hard- and soft-cores, see Fig.~\ref{3taurule}(b). This potential was introduced in Refs.~\cite{Ryzhov JCP2008,RyzhovPRE2009,Ryzhov JCP2011_1} in the following form:
\begin{gather}
  U(r)=\varepsilon\left(\frac{\sigma}{r}\right)^{n}+\frac{\varepsilon}2\left\{ 1-\tanh\left[k_0\left(r-\sigma_1 \right)\right]\right\}.
\end{gather}
Using the identity
\begin{gather}
  n_F(x)=\frac12\left\{1-\tanh\frac x2\right\},
\end{gather}
where the Fermi-function, $n_F(x)=1/[1+\exp(x)]$, we rewrote the potential of the collapsing soft spheres model:
\begin{gather}
  U(r)=\varepsilon\left(\frac{\sigma}{r}\right)^{n}+\varepsilon n_F\left[2k_0\left(r-\sigma_1 \right)\right].
\end{gather}

\subsection{Angell plot\label{Angell_plot}}
\textit{Angell plot} for temperature dependence of diffusion coefficient demonstrating essential deviation of $D(T)$ from Arrhenius law and so the fragile character of system dynamics, see Fig.~\ref{3taurule}(c). The construction of such plot is the traditional way to distinguish between the strong and fragile glassformer~\cite{Angell0}.

\section{Discussion}
\begin{figure}[b]
  \centering
  \includegraphics[height=7.105cm,width=7.09cm]{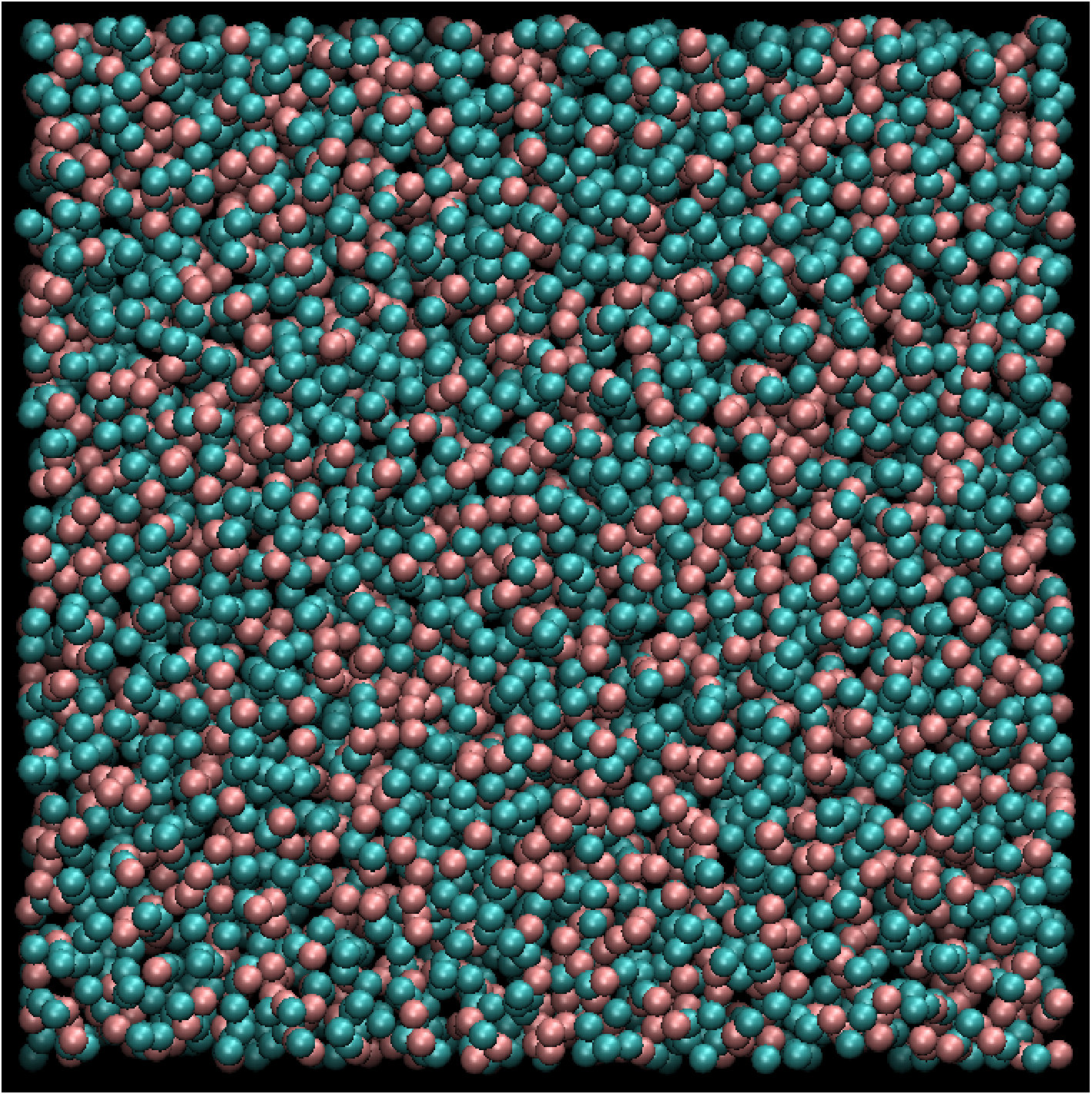}
  \\
  \caption{(Color online) Snapshot of the typical distribution of particles over the simulation volume. Here $T=0.1$, and $\rho=0.6$. We mark by red (blue) color particles that have a neighbor at distance~$\approx r_1$ ($r_2$) [$r_1$ and $r_2$ are defined in Fig.~5 of our paper].}\label{snapshot}
\end{figure}
What is the basis for the paradigm that no glass can be expected for purely repulsive potential for quasiequilibrium cooling? First of all this is ``experimental'' fact. Namely, it is definitely established last decades from numerical molecular-dynamic experiments that purely repulsive potential in one-component isotropic system does not allow to freeze the system into a glass while cooling is quasiequilibrium (as it was defined above): the system spontaneously crystalizes (there is lack of frustration built in this system…), see also discussion in Refs.~\cite{Kob1999,Molinero_Sastry_Angell,Bernu}. However, if the cooling rate is large enough, larger than the typical relaxation rate, then it is possible to freeze such system into a glass with $T_g$ essentially depending on the cooling rate.  But in this dynamical cooling regime the region above $T_g$, we focus on in our paper, is not accessible where glassy slowing down is already present but the system is still in the liquid state. The dynamical cooling is in fact the method to ``cook'' the glass and find $T_g(v)$.  Both regimes, qusiequilibrium and dynamical, are accessible experimentally. Therefore, if we want to study the mechanism of glassy slow down in supercooled state we have to use quasiequilibrium method.  So our model system should not spontaneously crystallize being supercooled in quasiequilibrium conditions due to some frustrations. If we will speak on the intuitive level it becomes difficult to imagine frustration in simple monatomic pure repulsive system: potential is isotropic and what is most important only one characteristic scale is built in.  So we propose the repulsive potential with additional source of frustration. Our potential has the soft step and the simple liquid with this interaction has two characteristic scales. The competition between these scales makes the system effectively quasibinary. And this is the origin of frustration in our system.

The physical origin of the non-Arrhenius behavior of fragile glass formers is an area of active investigation in glass physics. Advances over the last decade have linked this phenomenon with the presence of locally heterogeneous dynamics in fragile glass formers; i.e. the presence of distinct (if transient) slow and fast regions within the material~\cite{Stillinger,Angell2000}. The quasibinary character of our particle system allows to make the selection among the particles: we mark by red (blue) color particles that have a neighbor at distance of the order of the first (second) maximum in rdf [hard-core and soft-core diameters], see Fig.~5 in the main paper. To demonstrate the high degree of the heterogeneity in our system we place the snapshot of the spatial particle distribution over the simulation volume, see Fig.~\ref{snapshot}.  The high spatial heterogeneity of our system favors the existence of the locally heterogeneous dynamics within the system because of different free volumes (and so diffusion coefficients) of particles with different effective radii.


\begin{thebibliography}{99}
\bibitem{footnote} It is known that $T_g$ is somewhere between approximations of $T_g$ obtained from (VF) and (MCT).
\end{thebibliography}
\end{document}